\documentstyle[preprint,aps]{revtex}
\begin{document}
\newcommand{\bfsigma}{\mbox{\boldmath $\sigma$}}
\newcommand{\bbti}{$^{48}$Ti}
\newcommand{\bbca}{$^{48}$Ca}
\newcommand{\bbsc}{$^{48}$Sc}
\draft
\title{Gamow-Teller strength distributions in $fp$-shell nuclei}
\author{P. B. Radha${}^1$\cite{radhaadd}, 
D. J. Dean${}^2$, S. E. Koonin${}^1$, 
K. Langanke${}^3$, P. Vogel${}^4$}
\address{${}^1$W. K. Kellogg Radiation Laboratory, California
Institute of Technology,\\ Pasadena, California 91125, USA\\
${}^2$ Physics Division, Oak Ridge National Laboratory, P.O. Box 2008,
\\Oak Ridge, Tennessee 37381, USA\\
${}^3$ Institut for Physics and Astronomy, University of Aarhus,\\
DK-8000 Aarhus C, Denmark \\ 
${}^4$ Division of Physics, Mathematics and Astronomy,
California Institute of Technology,\\ Pasadena, California 91125, USA}

\date{\today}
\maketitle

\begin{abstract}
We use the shell model Monte Carlo method to calculate
complete $0f1p$-shell response functions for Gamow-Teller (GT)
operators and obtain the corresponding strength distributions
using a Maximum Entropy technique. The approach is validated
against direct diagonalization for {\bbti}. Calculated  
GT strength distributions agree well with data from 
$(n,p)$ and $(p,n)$ reactions  
for nuclei with $A=48-64$.   
We also calculate the temperature 
evolution of the GT$_+$ distributions for representative nuclei and 
find that the GT$_+$ distributions  
broaden and the centroids shift to lower 
energies with increasing temperature.  
 
\end{abstract}


\narrowtext

\section{Introduction}

The Gamow-Teller (GT) properties of nuclei in the medium mass
region of the periodic
table are crucial determinants of
the precollapse evolution of a supernova~\cite{bethe:1992}.
The core of a massive
star at the end of hydrostatic burning is stabilized by electron degeneracy
pressure as long as its mass does not exceed the appropriate Chandrasekhar
mass $M_{CH}$. If the core mass exceeds $M_{CH}$, electrons are captured by
nuclei. For many of the nuclei that determine the electron
capture rate in this early stage of the presupernova \cite{aufderheide:1994},
Gamow-Teller (GT) transitions contribute significantly. 
Due to insufficient experimental information, the GT$_+$
transition rates have so far been treated only qualitatively in
collapse simulations, assuming the GT$_+$ strength to reside in a single
resonance whose energy relative to the daughter ground state has been
parametrized phenomenologically \cite{FFN:1980}; 
the total GT$_+$ strength has
been taken from the single-particle model. However, recent $(n,p)$ experiments
\cite{williams:1995,alford:1990,vetterli:1989,elkateb:1994,ronnquist:1993},
show that the GT$_+$ strength is
fragmented over many states, and that the total strength is significantly
quenched compared to the single-particle model. (A recent update of the
GT$_+$ rates for use in supernova simulations assumed a constant quenching
factor of 2 \cite{aufderheide:1994}.) 

In this paper, we describe our calculations of
Gamow-Teller strength distributions in iron region nuclei: 
the shell model Monte Carlo
(SMMC) technique is used to obtain the response functions of the Gamow-Teller 
operators in the full $0\hbar\omega$ $fp$-shell model space. 
These response functions are related to the strength distributions
through an inverse Laplace transformation, which we carry out
using a Maximum Entropy method.

Our starting point is the 
interacting shell model~\cite{mayer:1955}, which gives an accurate and 
consistent description of the properties of 
light nuclei~\cite{cohen:1965,wildenthal:1988} when an appropriate interaction
is used. 
In the shell model, nucleons 
occupy a spectrum of single-particle orbitals that are formed by the
presence of an assumed mean field. These nucleons interact 
through a residual effective interaction,
which is derived from a realistic nucleon-nucleon potential
through the $G$-matrix formalism\cite{hj:1995}.
The resultant interaction matrix elements require some minimal tuning 
to optimally account for known spectroscopic properties. 
In the conventional approach, 
the solution to the shell model 
is obtained by diagonalizing the 
nuclear Hamiltonian in a suitably chosen basis of many-particle 
configurations. 
Since the 
Hamiltonian matrix to be diagonalized grows combinatorially with 
the size of the single-particle basis and the number of valence 
nucleons, realistic calculations are feasible in 
the full $fp$-shell only for nuclei with $A \le 50$.  
Hence, the traditional calculation of various nuclear properties 
for medium-heavy and heavy nuclei lies beyond the scope of 
direct-diagonalization methods except in a severely truncated model
space. 
 
The SMMC method
\cite{johnson:1992,lang:1993,ormand:1994,physrep:1997} 
scales more gently with the problem 
size than do traditional direct-diagonalization techniques, 
allowing larger, and 
hence more realistic, calculations.   
This method exploits the fact that most of the billions 
of configurations in nuclei are unimportant for general 
nuclear properties, so that only a subset of the relevant configurations 
needs to be sampled. 
Observables are calculated as thermal averages in a
canonical ensemble of nuclear configurations, so that 
nuclei at finite temperature can be studied quite naturally. 

SMMC methods were used in the first  
complete $0\hbar\omega$ calculations for a number of    
ground-state~\cite{alhassid:1994,dean:1994,karli:1995},
and finite-temperature properties~\cite{dean:1995} of mid-$fp$ shell nuclei.  
These studies used both the 
Richter-Brown~\cite{richter:1991} and the 
KB3~\cite{kb3:1981} residual interactions.
For the purposes of investigating 
Gamow-Teller transitions, the KB3 interaction
(obtained by minimally modifying the monopole strength in the 
original Kuo-Brown matrix elements~\cite{kbrown:1968})
is well-suited for full $0\hbar\omega$ studies throughout the lower-$fp$ shell
region~\cite{caurier:1994}. 
Observables that have been calculated with this interaction
in the SMMC approach include 
the energy $\langle H\rangle$,  
the total $B(E2)$, $B(M1)$, GT strengths, and 
various pairing properties; the calculated
ground-state properties compare very well with experiment.  
Importantly, these studies showed that the experimentally observed
quenching of the total GT strength is consistently reproduced by the
correlations within the full $fp$-shell if a renormalization of the spin
operator by the factor 0.8 is invoked \cite{karli:1995,caurier:1994}.
The same
renormalization factor had already been deduced from sd-shell 
\cite{wildenthal:1988} and
$fp$-shell nuclei with $A \leq 49~$\cite{caurier2:1996,poves:1997} 
and thus appears to be universal. 

In Section II, we reveiw the SMMC method and its application 
to response   
functions. We apply a Maximum Entropy (MaxEnt) method
to perform the required inverse 
Laplace transform of the SMMC response functions;  
our implementation of MaxEnt for SMMC
is discussed in 
Section III. Section IV includes a validation of these methods 
against direct diagonalization for GT transitions in {\bbti},
and we present GT strength functions for several
heavier nuclei in the $fp$-shell ($A=48-64$) 
where experimental data are available. 
We also discuss the evolution of these distributions with  
temperature. A brief conclusion follows in Section V.   

\section{The Shell Model Monte Carlo Method} \label{smmc}

The SMMC method is based on a  
statistical formulation of the 
nuclear many-body problem. In the finite-temperature 
version of this approach, an observable is
calculated as the canonical expectation value  
of a corresponding operator $\hat{\cal A}$ at a given temperature $T$ 
and is given by~\cite{johnson:1992,lang:1993,ormand:1994,physrep:1997} 
\begin{equation}
\langle \hat {\cal A}\rangle=
{{\rm Tr_A} [\hat {\cal A} e^{-\beta \hat H}]\over
{\rm Tr_A} [e^{-\beta \hat H}]},
\end{equation}
where $\hat U=\exp(-\beta \hat H)$ is the imaginary-time 
many-body propagator,
${\rm Tr_A} \hat U$ is the canonical partition function
for $A$ nucleons, $\hat H$ is the shell model Hamiltonian,
and $\beta=1/T$ is the inverse temperature.  

In terms of a spectral expansion, the total strength of a transition 
operator $\hat {\cal A}$
is then given by the following expectation value:   
\begin{equation}
B({\cal A}) \equiv \langle \hat {\cal A}^\dagger \hat {\cal A}\rangle =
{{\sum_{i,f} e^{-\beta E_i} \vert \langle 
f \vert \hat {\cal A} \vert i \rangle \vert ^2} \over {\sum_i e^{-\beta
E_i}}}, 
\end{equation}
where $\vert i\rangle$ ($\vert f\rangle$) are the many-body states of the 
initial (final)
nucleus with energy $E_i$ ($E_f$). The total strength from the ground state
can be obtained by choosing a sufficiently large value for $\beta$ such
that only
the ground state contributes due to the Boltzmann weight.   

In addition to the ``static'' strength [Eq. (2)], one can calculate 
for an imaginary-time $\tau$, 
the response function, $R_{\cal A}(\tau)$, which describes
dynamical behavior and contains information about the nuclear spectrum:  
\begin{equation}
\label{response function}
R_{\cal A}(\tau)\equiv\langle \hat {\cal A}^\dagger(\tau) \hat {\cal
A}(0)\rangle = 
{ {\rm Tr}_A[e^{-(\beta-
\tau) \hat H} \hat {\cal A}^\dagger e^{-\tau \hat H} \hat {\cal A}] 
\over {\rm Tr}_A [e^{-\beta \hat H}]}
={\sum_{if} e^{-\beta E_i} e^{-\tau(E_f-E_i)} {\vert \langle f \vert 
\hat{\cal A}  
\vert i \rangle \vert}^2 \over \sum_i e^{-\beta E_i} }.
\end{equation}
The strength distribution     
\begin{equation}
S_{\cal A}(E)={{\sum_{if} \delta(E-E_f+E_i) 
e^{-\beta E_i} \vert \langle f \vert 
\hat {\cal A} \vert i \rangle \vert ^2} \over {\sum_i e^{-\beta E_i}}}
\end{equation}
is related to
$R_{\cal A}(\tau)$  
by a Laplace Transform,
\begin{equation}
\label{laplace transform}
R_{\cal A}(\tau) = \int_{-\infty}^\infty S_{\cal A} (E) e^{-\tau E} dE.
\end{equation}

Note from Eq.~(3) that
ground-state to ground-state transitions require large
$(\beta-\tau)$ in addition to large $\beta$.  
The large-$\tau$ behavior of $R_{\cal A}$ allows, in principle, 
a measurement of the specific transition between the ground state 
and the lowest allowed final state 
by the operator; the slope of log$_e$[$R(\tau)$] in this 
limit provides the transition energy, and the 
intercept measures the transition strength.     

The SMMC canonical expectation values are based on
the discretization of the many-body propagator, $e^{-\beta H}$,
into a finite number of ``time'' slices, $N_t$, each of duration 
$\Delta\beta=\beta/N_t$. At each time slice the many-body propagator 
is linearized via the Hubbard-Stratonovich 
transformation~\cite{hubbard:1957,strato:1957}; 
observables are thus expressed as path integrals of one-body 
propagators in fluctuating auxiliary fields. The integration 
is carried out by a Metropolis random walk~\cite{Met:1953}.

To circumvent the ``sign problem'' encountered in the SMMC 
calculations with realistic interactions,
we use the extrapolation procedure 
outlined in Refs.~\cite{alhassid:1994,dean:1995}. 
Yet another, but distinct, source of the sign problem 
is an odd number of nucleons  
in the canonical expectation values~\cite{physrep:1997}.  
We overcome this problem by a number-projection technique, 
first employed in
\cite{dean:1994} and subsequently used in
\cite{physrep:1997}, that allows us to extract information concerning
odd-$A$ nuclei from the neighboring even-even system. 

\section{The method of Maximum Entropy} \label{maxent}

Once we have the Gamow-Teller response functions, they must be inverted
to obtain strength distributions. 
The inverse of the Laplace transform (5) required to extract the strength
functions is 
an ill-conditioned numerical problem~\cite{press:1992}. The kernel (which in 
this case is $e^{-\tau E}$) acts as a smoothing  operator and thus 
the solution, for which the kernel must be inverted, will be extremely
sensitive
to small changes, ({\it i.e.}, to errors) in the input data. In
this section, we describe a Maximum Entropy procedure to carry out
the inversion~\cite{physrep:1997}.  

Consider the $\chi^2$-deviation of the data, $r_i \equiv
R(\tau=i\Delta\beta)$,
with errors, $\sigma_i$, 
from the fit values $F_i\{S\}$ produced by the trial inverse and 
obtained according to 
Eq.~(\ref{laplace transform}),
\begin{equation}
\label{chisq}
\chi^2\{S\}= \sum_i {({r_i - F_i\{S\} \over \sigma_i})}^2. 
\end{equation}
Direct minimization of $\chi^2$ is numerically stable only in the
simplest of circumstances (such as few-parameter data fitting). 
Combining $\chi^2$ with some other
auxiliary well-conditioned functional, $P\left\{S\right\}$, such 
that $P\{S\}$ has a minimum at the smooth solution, $S(E)$, and  
penalizes strongly oscillating functions, leads to a compromise 
between fitting the data and the expected smoothness of the inverse.
Thus one minimizes the joint
functional
\begin{equation}
\label{functional}
{{1}\over{2}}\chi^2\left\{S\right\}+P\left\{S\right\} \; .\end{equation}

The functional $P\{S\}$ is chosen as the information theoretic entropy,
\begin{equation}
\label{pofs}
P\left\{S\right\}=\alpha \int dE \left[ m(E)-S(E)+
S(E)\ln\left({{S(E)}\over{m(E)}}\right)\right]\;,
\end{equation}
where $m(E)$ is a default model and $\alpha$ is an adjustable 
parameter that both specify the 
{\it a priori} knowledge of $S(E)$. 

In order to minimize the functional (\ref{functional}),
we employ the technique
of Ref.~\cite{meshkov:1994}, which involves an iterative 
sequence of linear programming problems. We first
expand Eq.~(\ref{pofs}) to second order in $S(E)$ about some positive function
$f(E)$ to obtain
\begin{equation}
\label{expansion of pofs}
P\left\{f\mid S\right\}=\alpha \int dE \left\{\left(m-{{f}\over{2}}\right)
+\left[\ln\left({{f}\over{m}}\right)-1\right]S
+{{S^2}\over{2f}}\right\} \; .
\end{equation}
If the true minimum $S(E)$ of the non-quadratic functional in
Eq.~(\ref{pofs}) is taken as a point of expansion of $f(E)$ in
[Eq.~(\ref{expansion of pofs})], then
it also gives the minimum of the corresponding quadratic functional
\begin{equation}
S(E)=\min_a \left[{{1}\over{2}}\chi^2\left\{a\right\} + 
P\left\{S\mid a\right\}\right] \; . \end{equation}

Since we require extraction of positive strength function, we iterate 
while retaining partially the result of the previous iteration as,
\begin{equation}
S^{(n+1)}=\min_{S\ge 0}\left[{{1}\over{2}}\chi^2\left\{S\right\}+
P\left\{f^{(n)}\mid S\right\}\right]\;, \end{equation}
with
\begin{equation}
f^{(n)}(E)=\xi S^{(n-1)}(E)+(1-\xi)S^{(n)}(E)\;, \end{equation}
and the default model as the starting approximation to $S$,
\begin{equation}
S^{(0)}(E)=S^{(-1)}(E) \equiv m(E)\;. \end{equation}
The rate of convergence and stability are controlled by the 
mixing parameter $0 < \xi < 1$; 
a value of $\xi=0.3$ is a reasonable choice
to guarantee stability. Typically, convergence to the ``true'' solution
is obtained in less than 40 iterations. 
In this way, the minimization of a general functional that is intrinsic 
to a Maximum Entropy approach is reduced to an iterative procedure 
in which each step requires the minimization of a quadratic functional 
with linear inequality constraints.  

Some general remarks regarding this inversion technique are called for.   
Since $R(\tau)$ is calculated 
at discrete values of imaginary time and, in principle, up to an imaginary 
time $\beta$, the smallest energy that 
can be resolved in $S(E)$ is of order $1/\beta$, and the largest 
is the inverse of the discretization size, $1/\Delta\beta$. In practice, 
numerical noise forces a cut-off in the largest $\tau$ value that can 
be used, thus decreasing the energy resolution.

As we mentioned above, the default model can be chosen by investigating
the characteristics of the response function. 
From Eq.~(3), one sees that 
$d\ln[R(\tau)]/d\tau \vert_{\tau=0}$ 
gives the centroid of the 
distribution in the parent nucleus, and thus 
in the case of the GT$_+$ operator 
we choose for the default model 
a Gaussian 
with a peak at this energy and with a width of $1.5-2$ MeV;
this width can be estimated from $d^2\ln[R(\tau)]/d\tau^2 \vert_{\tau=0}$.
The parameter $\alpha$ is the inverse of the total strength of the
distribution, and is calculated from the default model as
$\alpha=\left[\int dE m(E)\right]^{-1}$.
In the case of the GT$_-$ operator, we make  
a better guess for the default model by including some features     
of the distribution. Experimental distributions typically have 
three regions: the $T=T_z$, and $T=T_z+1$ 
regions distributed around 6 MeV   and 12 MeV, respectively, 
and a more fragmented region at lower energies.   
We choose for our GT$_-$ default model two gaussians
with the same widths, each centered at the appropriate energy. 
The lower energy part of the distributions is governed by the high
$\tau$ region of the response function. Although this
region of the response function
is sometimes contaminated by large statistical fluctuations,
the reconstruction tends to give a low-energy peak that well 
describes these more discrete transitions. 

\section{Gamow-Teller strength distributions}
The GT operators are defined as 
${\bf GT}_{\pm}=\sum_l \bfsigma_l\tau^{\pm}_l$, where
$\bfsigma_l$ is the Pauli spin operator
for nucleon $\it l$ and
$\tau^-_l$ ($\tau^+_l$) is the isospin lowering (raising) 
operator that changes a neutron (proton) into a proton (neutron);  
they thus describe charge-changing decay modes.   
GT strength distributions play an important role in two 
very different contexts. In the astrophysical context, medium-heavy nuclei 
at a finite temperature in the core of a pre-supernova 
capture electrons. A strong phase space dependence makes the relevant  
electron capture rates more sensitive to GT {\it distributions}
than to total strengths~\cite{aufderheide1:1993,aufderheide2:1993} and  
thus necessitates complete $0\hbar\omega$    
calculations of these distributions.  
GT strengths are also
important in studies of double beta decay~\cite{boehm:1987}. The two-neutrino 
mode of this decay, which 
provides important
confidence in extracting the neutrino mass from zero-neutrino decay 
experiments, is equivalent to a description of the GT
strength functions from the ground states of the parent and daughter
nuclei. 
Thus, any reliable calculation of the two-neutrino matrix element
must accurately describe these strength distributions. 

In the following 
sections we demonstrate and validate the MaxEnt method for the GT 
operator by comparing our results with direct diagonalization. We then 
compare our results with experimentally obtained distributions for various 
$fp$-shell nuclei. In what follows we will use   
the renormalized GT operator corresponding to 
${\bf GT_{\pm}}/1.26$~\cite{karli:1995,caurier:1994}.

\subsection{Comparison with direct diagonalization}
Direct-diagonalization results in the complete $fp$-shell 
can be obtained for nuclei with $A \leq 48$. 
We choose {\bbti} for a comparison     
and in Fig. 1, we show our results for this nucleus. 
The lower left-most panel shows the GT$_+$ 
response function, $R(\tau)$, for  {\bbti} as measured in the parent  
and the middle lower panel shows the extracted strength distribution, 
$S(E)$, in the daughter, {\bbsc}. 
Also shown in the same panel is the direct-diagonalization
result~\cite{caurier:1990}. 
The discrete transitions found in the direct diagonalization have been 
smeared with a gaussian of width 0.25~MeV in order to 
facilitate comparisons. 
While the SMMC total strength ({\it i.e.}, the area under the 
curve) $B(GT_+) = 0.72\pm0.11$~\cite{karli:1995} 
compares very well with the 
direct-diagonalization value of $0.79$~\cite{caurier:1994}, 
the SMMC can recover only 
gross features of this distribution.
In particular, the peak is somewhat too narrow,
mainly due to the information lost by the Laplace
transform. This attribution was checked by calculating the response
function $R(\tau)$ for the
direct diagonalization 
distribution. (The peaks were smeared by Gaussians of 0.25 MeV width to
account for the SMMC finite discretization.)
This response function is shown in the lower left panel of Fig.~1, and 
agrees well with the SMMC result.

The lower right-most panel in Fig. 1 shows the energy dependence of  
the cumulative strength, $\int_0^{E^\star} S(E^\prime) dE^\prime$, where
$E^\star$ is the excitation energy in the daughter. One 
can see that the SMMC recovers the centroid and the width of the 
distribution reasonably. 

A brief remark about the possible sources 
of error is in order. Since our MaxEnt procedure 
provides a most probable 
extraction of the strength function, the strength distributions do not have 
error bars associated with them. However, from the SMMC error bars 
for $R(\tau)$, we estimate the error in the position of the centroid 
to be about $0.5$ MeV. In addition, we note that  
the response functions are measured in 
the parent nucleus, and to obtain the energy in the daughter we use 
the experimental mass excesses and a parametrization of the Coulomb 
energy as defined in~\cite{caurier:1994}. [In the
test case ($^{48}$Ti), we exactly calculate this mass difference.]
This parametrization 
provides a good overall description of the masses 
of the nuclei in this region~\cite{karli:1995}.  
We find an average deviation between $0.1$ MeV 
(for $A=48$ nuclei) and $0.5$ MeV (for $A=54$ nuclei) of our 
calculated binding energies from experimental values, 
suggesting that our procedure is quite justified.
 
The upper panels of Fig. 1 show our results for the GT$_-$ operator in  
{\bbti}. The total strength, $B(GT_-)$, can be readily obtained 
from the renormalized Ikeda sum rule, $B(GT_-)-B(GT_+)=3(N-Z)/(1.26)^2 $ 
which is obeyed 
by both the SMMC and direct-diagonalization calculations. 
The GT$_-$ operator takes the $N>Z$ parent nucleus (with $T=T_z+1$) 
to $T=T_z$(dotted),
$T=T_z+1$ (dashed), and $T=T_z+2$ (not shown) states in
the $^{48}$V daughter. The $T=T_z$ states
are the lowest in energy and  contain most (85\% in this case) of
the strength. Assuming in the default model 
that the centroid
of the $T=T_z+1$ states is located 5 MeV higher than the centroid of
the $T=T_z$ states, we obtain a good reproduction of both
components of the
strength distribution. This general assumption 
is experimentally valid in the even-even nuclei in this region. 
We also see at low energy
a hint of the discrete low-energy states in the reconstruction. 

\subsection{Comparison with experiment}\label{comparison with experiment}

Experimental GT distributions  are obtained from   
intermediate-energy charge exchange $(n,p)$ [or $(p,n)$] cross sections 
at forward angles, which are proportional to  
the GT strength~\cite{bertsch:1987}.  
These experimental 
distributions typically extend only to $8$ MeV in the daughter nucleus 
to exclude contributions from other multipolarities.  

We first compare our {\bbti} result for the GT$_+$ distribution against 
experiment, as shown in Fig.~2. 
To simulate the finite experimental resolution 
and presentation of the data, the SMMC results have been 
smeared with Gaussians of standard 
deviation of $1.77$ MeV, following Ref. ~\cite{caurier:1995}. Our
results are represented by the dotted line in Fig.~2, while the 
diagonalization
results are shown as a solid histogram. The smeared diagonalization result
is shown by the dashed line in the figure. 
The experimental $B(GT_+)$ distribution, shown as solid dots, 
sums $1.42\pm0.2$ \cite{alford:1990}
compared to our 
renormalized value of $0.71\pm0.11$. We find that the calculated 
$0\hbar\omega$ GT$_+$ strength extends only over the region 
E$^\star < 8$ MeV 
(in agreement with the experimental value for this range of energy 
$B(GT_+)$ = 0.77 $\pm$ 0.1). 
This suggests that 
the GT$_+$ strength observed for E$^\star > 8$ MeV 
corresponds to correlations 
outside our $0\hbar\omega$ model space. A similar conclusion has been 
reached in Ref.~\cite{caurier:1994}.    
We note that the inadequacy of a $0\hbar\omega$ model space to 
describe the GT$_+$ distribution at E$^\star > 8$ MeV 
might have some relevance to 
the $\beta\beta$ decay of {\bbca}~\cite{poves:1995},
where considerable $2\nu\beta\beta$
strength could be obtained  
from the overlap of this distribution with that of {\bbca} in the $(p,n)$ 
direction for these energies.                             
However, the measured $2\nu\beta\beta$ decay rate of $^{48}$Ca
\cite{balysh:1997} agrees
well with the calculation based on the $0\hbar\omega$ shell model, which
includes the $1/1.26$ normalization of the GT transition operator.

We now turn to a comparison of SMMC results with experiment for 
nuclei in the mid-$fp$-shell 
where complete direct-diagonalization calculations are not possible.
We first consider the $(n,p)$ reaction and  
in Fig.~3 we show our results for all even-even nuclei with $A=48-64$ 
for which data are available
~\cite{williams:1995,vetterli:1989,elkateb:1994}. 
The SMMC results have been smeared with Gaussians of standard 
deviation of $1.77$ MeV to account for
the finite experimental resolution, following Ref. ~\cite{caurier:1995}. 
Experimentally, the GT$_+$ strength is significantly 
fragmented over many states; the centroids and the widths of 
these distributions are reproduced very well in the SMMC approach. 
Our results for the total strengths are given in Table I. 
They agree with the data very well except for $^{64}$Ni where 
our calculation  underestimates the total experimental 
strength~\cite{williams:1995} 
suggesting the need to augment the model space with the 
$g_{9/2}$ and the $g_{7/2}$
orbitals. This shortcoming of the present model space is also visible    
in the GT$_+$ distribution, which places the GT$_+$  
peak approximately 1.5 MeV above the experimental peak and misses 
the second peak at E$\approx 6$ MeV, which is possibly due to 
$(g_{9/2}-g_{7/2})$ transitions.    

SMMC results for odd-$A$ nuclei in the $(n,p)$ direction are 
shown in Fig.~4, where again the centroids and widths of the  
distributions are in good agreement with the 
data~\cite{elkateb:1994,rapaport:1984,alford:1993}.  
Calculations for odd-$A$ nuclei are performed 
at a finite temperature of $0.8$ MeV. 
(The temperature dependence of 
these distributions will be discussed later in Section~\ref{distributions and
temperature}.) The response functions for the three nuclei in 
Fig.~4 are sampled from the partition functions of their 
neighbors, {\it i.e.}, 
$^{51}$V from $^{52}$Cr, $^{55}$Mn from $^{56}$Fe, and $^{59}$Co from 
$^{60}$Ni. The peaks of the observed GT$_+$ distributions in odd-$A$ 
nuclei in Fig.~4 are consistently at higher excitation energies in the 
daughter compared to the 
even-even cases in Fig.~3, a feature reproduced by the SMMC 
calculations. These higher excitation energies cause some $0\hbar\omega$ 
strength to lie above the typical 
8 MeV cut-off in odd-$A$ nuclei.
The data for $^{51}$V and $^{59}$Co have been analyzed
for additional strength above 8 MeV 
~\cite{aufderheide1:1993,aufderheide2:1993} 
(see Table 1), while, to our knowledge,
$^{55}$Mn has not been reanalyzed for potential GT strength at 
E$^\star >8$ MeV. 
For even-even nuclei the $0\hbar\omega$ GT$_+$ strength appears to 
be exhausted at energies below 8 MeV, in agreement with the SMMC 
results shown in Fig.~3. Our  
results for $^{51}$V 
and $^{55}$Mn show some strength above 8 MeV, but this
is not the case for $^{59}$Co.
     
In Fig.~5 we compare the GT$_-$ distributions for a few nuclei 
where experimental data are available~\cite{vetterli:1989,rapaport:1983}; 
the experimental data for 
$^{56}$Fe have been obtained from Ref.~\cite{caurier:1995}.  
From the cumulative strengths in the
right panels of Fig.~5, 
we can conclude that the 
SMMC approach reproduces the experimental distribution moderately well  
for the cases of $^{54}$Fe and $^{56}$Fe. For the Ni isotopes, 
only partial information is available about these distributions. 
For $^{58}$Ni 
the peaks in the experimental data~\cite{rapaport:1983} 
shown are to be associated with 
a finite width 1.3 MeV, 0.7 MeV, and 0.5 MeV for the peaks 
at 9.2 MeV, 11.2 MeV, and 13.0 MeV, respectively. The strength
in the giant resonance region 
between 6.4 MeV and 13.0 MeV is quoted as 
5.5, 
while we obtain $6.1$, which is consistent with the
uncertainty in the excitation energy. 
For $^{60}$Ni the experimental value of the total 
GT$_+$ strength~\cite{rapaport:1983}
is $7.2\pm1.8$ whereas we obtain $10.87\pm0.23$. As our 
calculation obeys the renormalized Ikeda sum-rule 
and reproduces the measured GT$_+$ strength,
the lower  
experimental value indicates some strength outside the 
experimental window of E$^\star > 14$ MeV. We also note that 
while Ref.~\cite{rapaport:1983} quotes an integrated strength of   
$6.22$ between $4.0$ and $14.0$, MeV we obtain a value of $4.65$.

\subsection{Temperature dependence of GT strengths}
\label{distributions and temperature}

We now turn to the temperature evolution of GT$_+$ 
strength functions. Representative strength 
distributions for two nuclei, $^{59}$Co and $^{60}$Ni, at  
several temperatures are shown in Fig.~6. Both figures are plotted as
a function of $E$, the energy transfer to the parent nucleus.
We note that   
the restriction of the model space to only $fp$-shell 
renders our calculation quantitatively unreliable for even-even nuclei 
at $T \gtrsim 1.4$ MeV~\cite{dean:1995},
while for the odd-$A$ cases this temperature is likely even lower.  

With increasing temperature, three distinct effects occur that
influence the GT strength distributions. 
\begin{itemize}
\item The number of states contributing to the thermal ensemble
increases. Due to the pairing gap in even-even nuclei, this occurs at a
higher temperatures in even-even nuclei than in odd-A nuclei.
\item GT transitions which are Pauli blocked at low temperatures due to
closed neutron subshells (e.g. the $f_{7/2}$ orbital) can become
thermally unblocked as neutrons are moved to excited orbitals with
increasing temperature. Similarly, protons which are thermally excited to
higher orbitals can undergo allowed GT transitions.
\item The ground state in even-even nuclei is dominated by like-nucleon
pairing. As indicated by SMMC calculations, these pairs break at around
$T=1$ MeV. Thus at low temperatures, a GT$_+$ transition involves
breaking a proton pair associated with an extra energy of 1-2 MeV. This
``penalty energy'' is removed at higher temperatures in states of higher
excitation energy, in which the pair correlations are diminished.
\end{itemize}
As we will discuss in the following, these three effects allow for an
understanding of the temperature dependence of the GT$_+$ strength
distributions.

In the case of $^{59}$Co, with increasing temperature, the entire 
distribution shifts to lower excitation energies. 
The total strength decreases
and the width of the distribution increases 
marginally with increasing temperatures. (We have
checked that in the
high-$T$ limit, $B(GT_+)$  
rises to the single-particle value as expected.)
Due to the lack of pairing of the odd particle in an odd-A nucleus,
states of various spins are more quickly populated than in the
even-even systems. These states
then make transitions to daughter states by the GT operator.
Thus, a plethora of states is easily accessible at moderate temperatures,
and the required excitation energy in the daughter is lower.

For $^{60}$Ni, the peak in the strength 
distribution remains roughly constant with increasing temperature,
while the width increases with 
the appearance of low-lying strength due to transitions from the 
thermally occupied to the empty excited orbitals. 
Note also that the centroid of the distribution remains constant at the
low temperatures and then shifts to lower excitation at higher
temperatures. 
The near constancy of the peak position in $^{60}$Ni at low temperatures
supports the shifting assumption 
[attributed to D.~Brink in Ref.~\cite{aufderheide:1991}]
which states that the centroid corresponding to
each parent excited state is shifted upward in the daughter nucleus by
the energy of the parent state~\cite{aufderheide:1991}. This hypothesis
assumes that the internal configuration of the low-lying states is
roughly the same. With increasing temperature, however, states with
other internal configurations gain statistical weight, and in particular,
the pair correlations in these excited states decrease. SMMC
calculations indicate that pairs break around $T=1$ MeV in even-even
nuclei, allowing for a dramatic increase in thermally populated states
in the parent at and above this temperature.
For these excited states, no coherence energy has to be paid as penalty
to break a proton pair in the GT transition, and the peak in the GT
distribution moves to smaller energies. We also note that at
temperatures $T \leq 1.3$ MeV the thermal ensemble already includes the
lowest excited $T+1$ states allowing for transitions at $E=0$.
In contrast, these transitions are not observed in $^{59}$Co at the
temperatures considered here, since the $T+1$ states in
this nuclei are at higher excited energies due to the larger neutron
excess. We also observe a gradual decrease of the peak position with
temperature in accordance with the fact that no pairing gap has to be
overcome in odd-A nuclei.

\section{Summary and Conclusions}

As mentioned in the Introduction, 
electron capture on iron region nuclei 
plays an important role at the onset of core collapse in a
massive star. Under these conditions, nuclei have a finite temperature of
0.2$-$0.6 MeV. It is well known that for nuclei with an opened
$fp$-shell
neutron configuration, $GT_+$ transitions dominate the electron capture
rate, and a strong phase-space dependence makes the rate sensitive to the
full $GT_+$ distribution, rather than only to the total strength.
Unfortunately, the $GT_+$ strength is not experimentally accessible for
those nuclei of importance in the presupernova collapse. Thus, collapse
studies have to rely on theoretical estimates which, until recently, could
not be performed with great confidence. This has now changed. As SMMC
calculations reproduce the measured data from first principles without
nucleus-specific data fitting (which has been necessary in previous
studies), they are reliable enough to predict the $GT_+$ distributions
for those astrophysically important nuclei not experimentally
accessible. SMMC calculations for these nuclei are in progress.
 
In this paper, we have calculated response 
functions for the Gamow-Teller operators for 
several nuclei in the $fp$-shell. We use the KB3 interaction, which 
is well suited for $0\hbar\omega$ calculations. 
Using an implementation of the MaxEnt technique, we have 
then obtained the corresponding strength distributions. 

The extracted Gamow-Teller distributions compare very well 
with both direct-diagonalization calculations and the experimentally 
obtained distributions. We note that we invoke the 
standard renormalization factor 
of $1/1.26$ for the transition operator, in keeping with the observation 
in $sd-$ and $fp$-shell nuclei that complete $0\hbar\omega$ calculations
require this overall renormalization for agreement with experiment. 

We have also studied the effect of finite temperature on Gamow-Teller 
distributions and have demonstrated for sample nuclei that our 
calculations at $T=0.8$ MeV should be adequate to describe the 
distributions required to calculate electron capture rates 
for the pre-supernova problem\cite{aufderheide:1994}.    
Studies of the Gamow-Teller strengths 
and electron capture rates for nuclei relevant 
to the presupernova collapse will be described elsewhere.

\acknowledgements

We acknowledge support from the
U.S. National Science Foundation
under Grants PHY94-12818 and PHY94-20470.
Oak Ridge National Laboratory is managed by Lockheed Martin Energy
Research Corp. for the U.S. Department of Energy under contract number
DE-AC05-96OR22464.  D.J.D acknowledges an E.~P.~Wigner Fellowship from ORNL.
KL has been partly supported by the Danish Research Council.
Grants of computational resources were provided by the Center
for Advanced Computational Research at Caltech and the
Center of Computational Science at ORNL.

\begin{table}
\caption
{Renormalized $B(GT_+)$ strengths as calculated in the
SMMC approach compared to experimental strengths
\protect\cite{williams:1995,elkateb:1994,rapaport:1984,alford:1993}.
The superscripts on the experimental results indicate the
upper limit of energies used to obtain the total strength. 
a: up to 14 MeV.
b: up to 12.5 MeV. c: up to 10 MeV. d: up to 8.5 MeV e: up to 8 MeV
}
\begin{center}
\begin{math}
\begin{array}{|c|c|c|} 
\hline
{\rm nucleus} & B(GT_+) {\rm (SMMC)} & B(GT_+) {\rm (expt)}  \\ \hline
^{48}{\rm Ti}  & 0.71\pm0.11& 1.31\pm0.2^a \\
^{51}{\rm V}  & 1.40\pm0.14& 1.48\pm0.03^b \\
^{54}{\rm Fe}  & 3.84\pm0.28& 3..1\pm0.6^c \\  
^{55}{\rm Mn}  & 1.84\pm0.36&  1.7\pm0.2^d\\
^{56}{\rm Fe}  & 2.51\pm0.17& 2.9\pm0.3^d \\
^{58}{\rm Ni}  & 4.23\pm0.31& 3.8\pm0.4^d \\
^{59}{\rm Co}  & 2.60\pm0.31&  2.39\pm0.07^b \\
^{60}{\rm Ni}  & 3.26\pm0.25& 3.11\pm0.08^e \\
^{62}{\rm Ni}  & 2.16\pm0.25& 2.53\pm0.07^e \\
^{64}{\rm Ni}  & 1.09\pm0.18 & 1.72\pm0.09^e \\ 
\hline
\end{array}
\end{math}
\end{center}
\end{table}

\begin{figure}
\caption{Left-most panels show the GT$_-$ (upper) and GT$_+$ (lower)
response functions calculated through the SMMC. The middle panels 
show the corresponding strength function and direct-diagonalization 
results~\protect\cite{caurier:1994,caurier:1990} in the corresponding
daughter.
For the GT$_-$ we show both the $T=T_z$ and $T=T_z+1$ channels, while
the dash-dot line in the GT$_+$ distribution comes from folding the
SMMC results with a gaussian corresponding to the experimental sensitivity.  
The right-most panels show the cumulative strengths as a function 
of the daughter excitation energy. For the GT$_-$ we show the cumulative
$T=T_z+1$ strength starting from the total in the $T=T_z$ channel.}
\label{fig1}
\end{figure}

\begin{figure}
\caption{Calculated strength function (smeared by the
experimental resolution) for the GT$_+$ operator for
$^{48}$Ti compared to the experimental data \protect\cite{alford:1990}.
Also shown is the shell model spectrum obtained by diagonalization, 
and smeared by 0.25 MeV (histogram) 
and by 1.77 MeV (dashed line) to account for the experimental
resolution.}
\label{fig2}
\end{figure}

\begin{figure}
\caption{Comparison of calculated GT$_+$ strength distribution against 
experiment~
\protect\cite{williams:1995,elkateb:1994,rapaport:1984,alford:1993} 
for even-even nuclei as function of excitation 
energy in the corresponding daughter nuclei.}
\label{fig3}
\end{figure}

\begin{figure}
\caption{Calculated GT$_+$ distributions for odd-$A$ nuclei.
Also shown are the experimental 
distributions~\protect\cite{elkateb:1994,rapaport:1984,alford:1993}. 
The energies are in the corresponding daughter.}
\label{fig4}
\end{figure}

\begin{figure}
\caption{Left panels: Calculated GT$_-$ distributions for several nuclei 
in the mid-$fp$ shell against distributions obtained from $(p,n)$ 
reactions~\protect\cite{vetterli:1989,rapaport:1983}. 
Right panels: Cumulative strength distributions 
versus daughter excitation energy for SMMC calculations and experiment.
$B(GT_-)$ from SMMC (solid circles) and from experiment (open circles) 
are shown staggered for clarity.}
\label{fig5}
\end{figure}

\begin{figure}
\caption{Temperature evolution of GT$_+$ strength distribution for 
sample nuclei (left: $^{59}$Co; right: $^{60}$Ni) 
versus parent excitation energy.}    
\label{fig6}
\end{figure}


\begin{references}

\bibitem[\dagger]{radhaadd}
Present address: Laboratory for Laser Energetics, University of Rochester,
250 E. River Road, Rochester, NY 14623 

\bibitem{bethe:1992}
H.A. Bethe, Rev. Mod. Phys. {\bf 64}, 491 (1992).

\bibitem{aufderheide:1994}
M.~B. Aufderheide, I. Fushiki, S.~E. Woosley, and D.~H. Hartmann, {Ap. J} {\bf
  S91},  389  (1994).

\bibitem{FFN:1980}
G.M. Fuller, W.A. Fowler, and M.J. Newman, 
Ap. J. {\bf S42}, 447 (1980); {\bf 48}, 279 (1982); 
Ap. J. {\bf 252}, 715 (1982); Ap. J. {\bf 293}, 1 (1985). 

\bibitem{williams:1995}
A.L. Williams et al, Phys. Rev. {\bf C51}, 1144 (1995).

\bibitem{alford:1990} 
W.P. Alford {\it et al.}, Nucl. Phys. {\bf A514}, 49 (1990).

\bibitem{vetterli:1989} 
M.C. Vetterli {\it et al.}, Phys. Rev. {\bf C40}, 559 (1989).

\bibitem{elkateb:1994} 
S. El-Kateb {\it et al.}, Phys. Rev. {\bf C49}, 3129 (1994).

\bibitem{ronnquist:1993} 
T. R\"onnquist {\it et al.}, Nucl. Phys. {\bf A563}, 225 (1993).

\bibitem{mayer:1955}
M. Mayer and J. Jensen, {\em {Elementary theory of nuclear shell structure}}
  (Wiley, New York, 1955).

\bibitem{cohen:1965}
S. Cohen and D. Kurath, Nucl. Phys. {\bf 73},  1  (1965).

\bibitem{wildenthal:1988}
B. Brown and B. Wildenthal, {Ann. Rev. Nucl. Part. Sci.} {\bf 38},  29
(1988).

\bibitem{hj:1995}
M. Hjorth-Jensen, T.T.S. Kuo, and E. Osnes, Phys. Reps. {\bf 261}, 125 (1995)
and references therein.

\bibitem{johnson:1992}
C.~W. Johnson, S.~E. Koonin, G.~H. Lang, and W.~E. Ormand, {Phys. Rev.. Lett.}
{\bf 69},  3157  (1992).

\bibitem{lang:1993}
G.~H. Lang, C.~W. Johnson, S.~E. Koonin, and W.~E. Ormand, {Phys. Rev. C} {\bf
  48},  1518  (1993).

\bibitem{ormand:1994}
W.~E. Ormand, D.~J. Dean, C.~W. Johnson, G.~H. Lang, and S.~E. Koonin, {Phys.
  Rev. C} {\bf C49},  1422  (1994).

\bibitem{physrep:1997}
S.~E. Koonin, D.~J. Dean, and K. Langanke, {Phys. Rep.} {\bf 278}, 1 (1997).

\bibitem{alhassid:1994}
Y. Alhassid, D.~J. Dean, S.~E. Koonin, G. Lang, and 
W.~E. Ormand, {Phys. Rev.  Lett.} {\bf 72},  613  (1994).

\bibitem{dean:1994}
D.J. Dean, B.P. Radha, K. Langanke, S.E. Koonin, Y. Alhassid, and
W.E. Ormand, Phys. Rev. Lett, {\bf 72}, 4066 (1994)

\bibitem{karli:1995}
K. Langanke, D.~J. Dean, P.~B. Radha, Y. Alhassid, and S.~E. Koonin, 
{Phys. Rev. C} {\bf 52},  718  (1995).

\bibitem{dean:1995}
D.~J. Dean, S.~E. Koonin, K. Langanke, P.~B. Radha, and Y. Alhassid, 
{Phys.  Rev. Lett} {\bf 74},  2909  (1995).

\bibitem{richter:1991} 
W. A. Richter, M. G. Vandermerwe, R. E. Julies, and B. A.
Brown, Nucl. Phys. {\bf A523} (1991) 325

\bibitem{kb3:1981}
A. Poves and A. Zuker, {Phys. Rep.} {\bf 70},  235  (1981).

\bibitem{caurier:1994}
E. Caurier, A. Zuker, A. Poves, and G. Martinez-Pinedo, {Phys. Rev. C} {\bf
  50},  225  (1994).

\bibitem{kbrown:1968}
T.~T.~S. Kuo and G.~E. Brown, {Nucl. Phys. A} {\bf 114},  241  (1968).

\bibitem{caurier2:1996}
G. Martinez-Pinedo, A. Poves, E. Caurier, and A.~P. Zuker, {Phys. Rev. C} 
{\bf 53},  R2602  (1996).

\bibitem{poves:1997}
G. Martinez-Pinedo, A.P. Zuker, A. Poves, and 
E. Caurier, Phys. Rev. {\bf C55}, 187 (1997).


\bibitem{hubbard:1957}
J. Hubbard, {Phys. Rev. Lett.} {\bf 3},  77  (1959).

\bibitem{strato:1957}
R. Stratonovich, {Dokl. Akad. Nauk. SSSR} {\bf 115},  1097  (1957).

\bibitem{Met:1953}
N. Metropolis, A. Rosenbluth, M. Rosenbluth, A. Teller, and E. Teller, 
J. Chem. Phys. {\bf 21}, 1087 (1953).


\bibitem{press:1992}
W.~H. Press, S.~A. Teukolsky, W.~T. Vetterling, and B.~P. Flannery, {\em
  {Numerical Recipes in FORTRAN }} (Cambridge University Press, Cambridge,
  1992).

\bibitem{meshkov:1994}
S. Meshkov and D. Berkov, {Int. J. of Mod. Phys.} {\bf C5},  987  (1994).

\bibitem{aufderheide1:1993}
M.~B. Aufderheide, S.~D. Bloom, D.~A. Ressler, and G.~J. Mathews, {Phys. Rev.
  C} {\bf 47},  2961  (1993).

\bibitem{aufderheide2:1993}
M.~B. Aufderheide, S.~D. Bloom, D.~A. Ressler, and G.~J. Mathews, {Phys. Rev.
  C} {\bf 48},  1677  (1993).

\bibitem{boehm:1987}
F. Boehm and P. Vogel, {\em {Physics of massive neutrinos}} (Cambridge
  University Press, Cambridge, New York, 1987).

\bibitem{caurier:1990}
E. Caurier, A. Poves, and A. Zuker, {Phys. Lett. B} {\bf 252},  13  (1990).

\bibitem{bertsch:1987}
G.~F. Bertsch and H. Esbensen, {Rep. Prog. Phys.} {\bf 50},  607  (1987).

\bibitem{caurier:1995}
E. Caurier, G. Martinez-Pindeo, A. Poves, and A.~P. Zuker, {Phys. Rev. C} {\bf
  52},  R1736  (1995).

\bibitem{poves:1995}
A. Poves, P.~B. Radha, K. Langanke, and P. Vogel, {Phys. Lett. B} {\bf
361},  1
   (1995).

\bibitem{balysh:1997}
A. Balysh et al. Phys. Rev. Lett. {\bf 77}, 5186 (1997).

\bibitem{rapaport:1984}
J. Rapaport {\it et~al.}, {Nucl. Phys. A} {\bf 427},  332  (1984).

\bibitem{alford:1993}
W.~P. Alford {\it et~al.}, {Phys. Rev. C} {\bf 48},  2818  (1993).

\bibitem{rapaport:1983}
J. Rapaport {\it et~al.}, {Nucl. Phys. A} {\bf 410},  371  (1983).

\bibitem{aufderheide:1991}
M.B.~Aufderheide, Nucl. Phys. A526, 161 (1991)

\end{references}
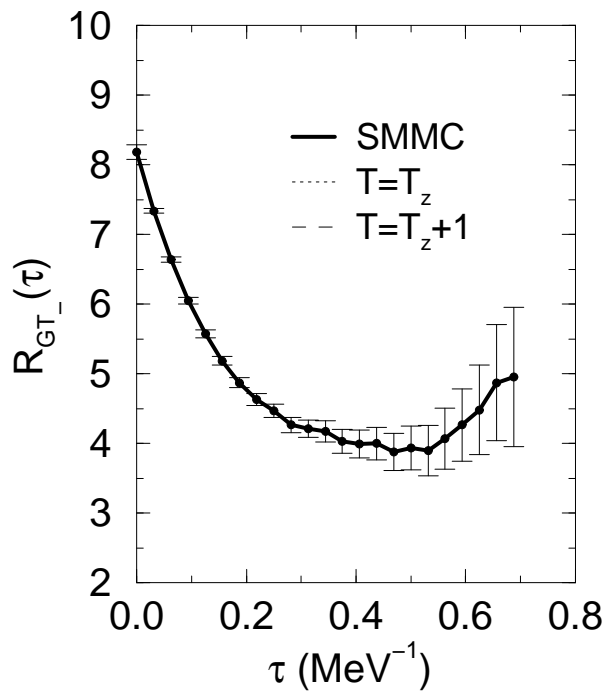
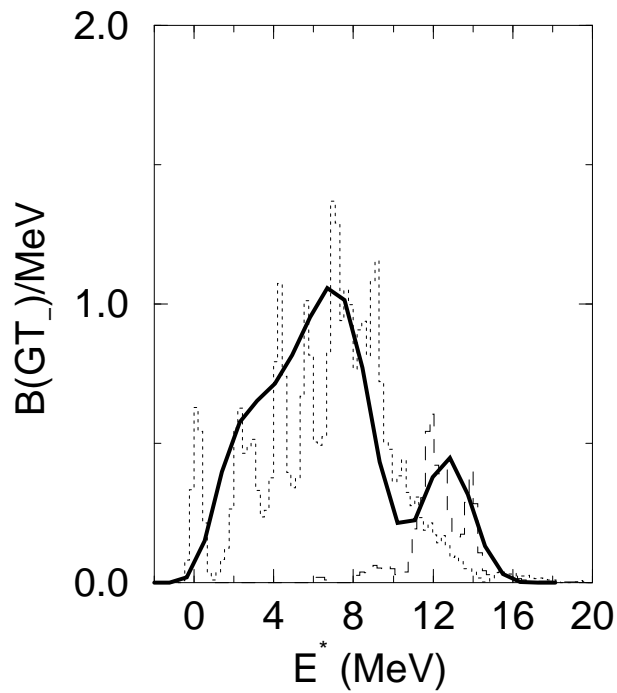
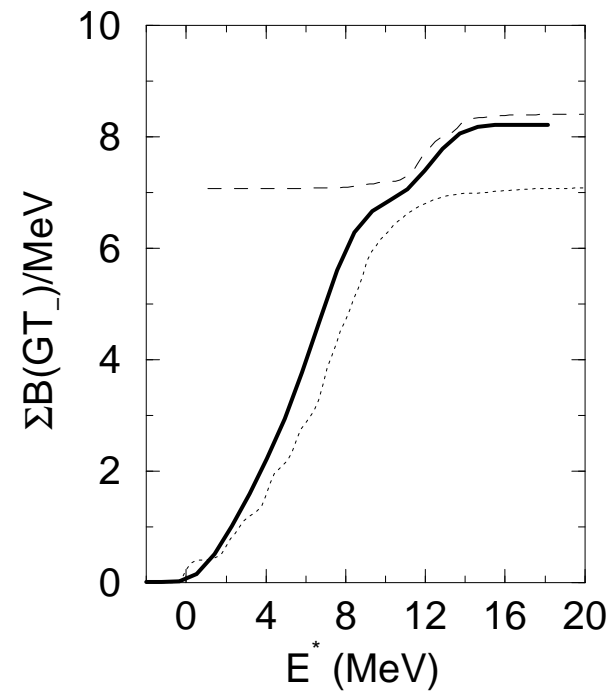
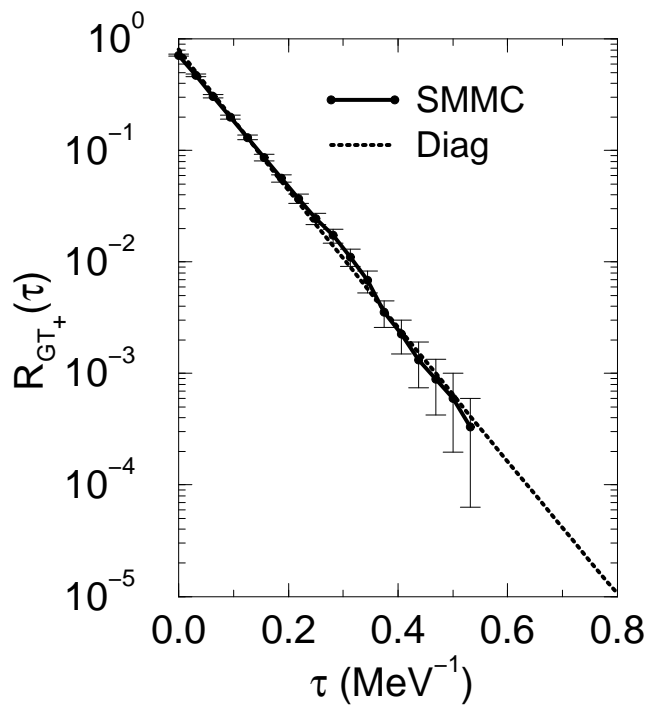
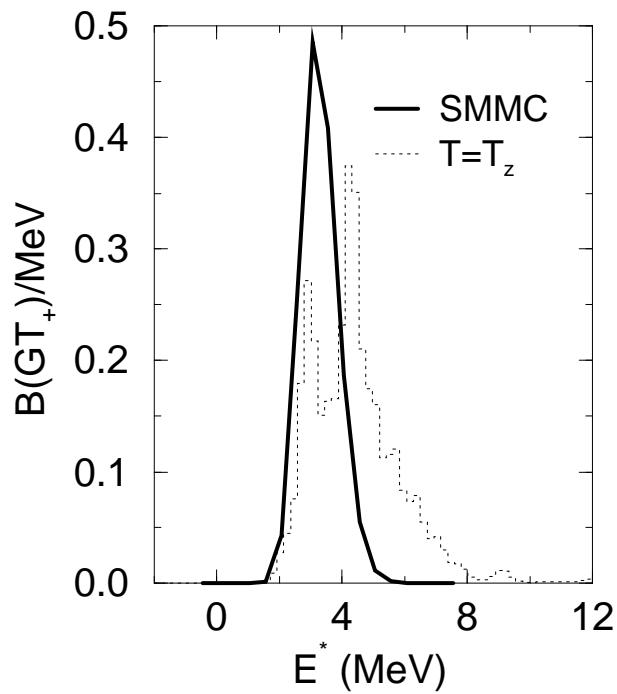
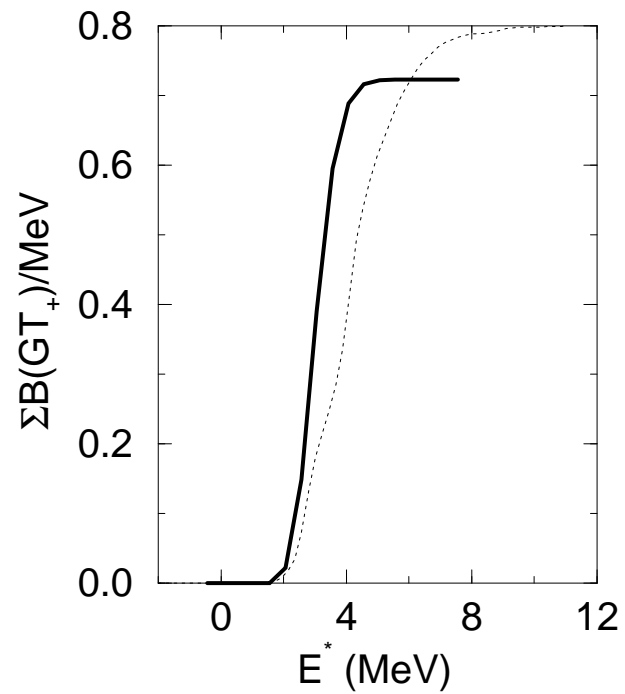
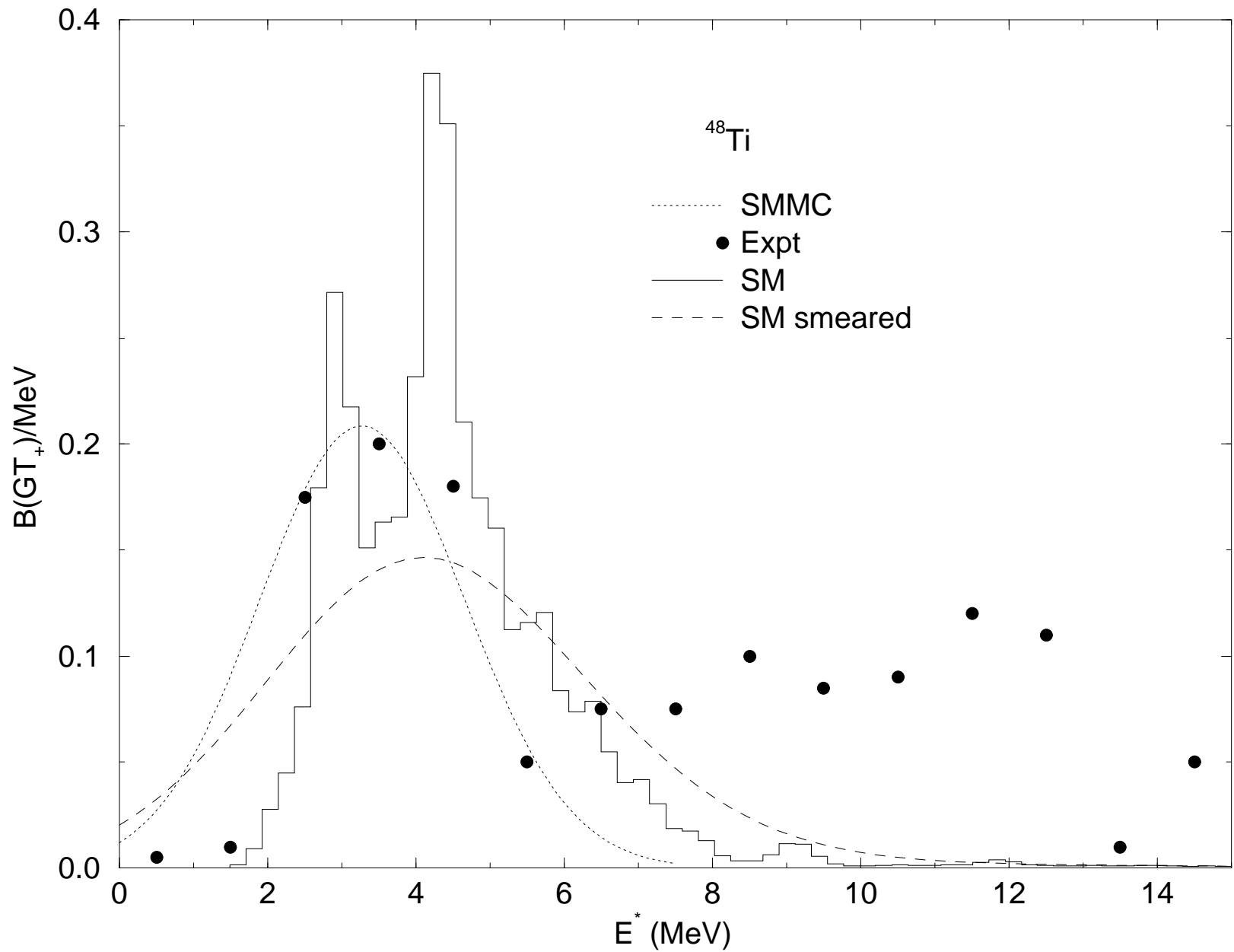
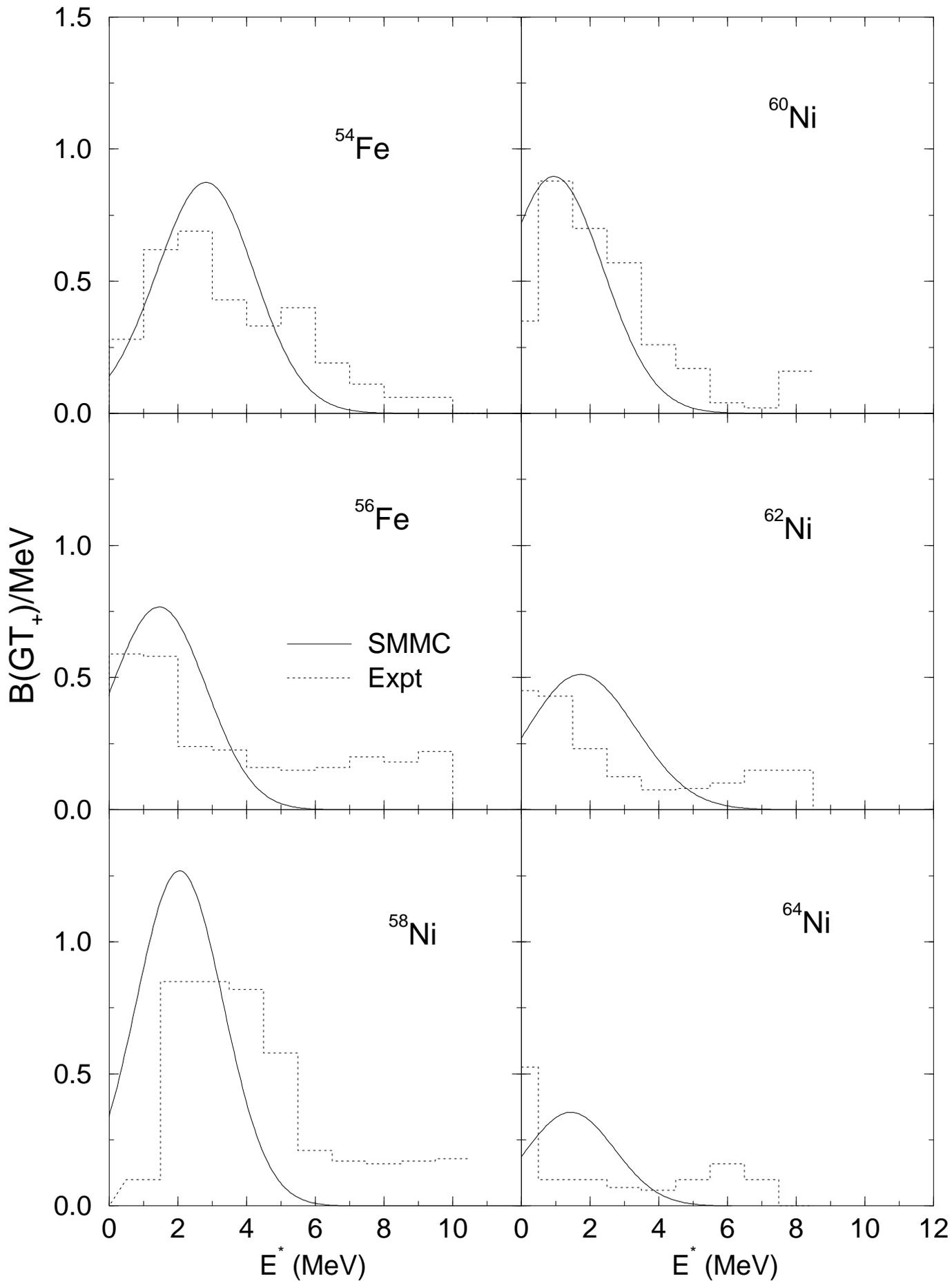
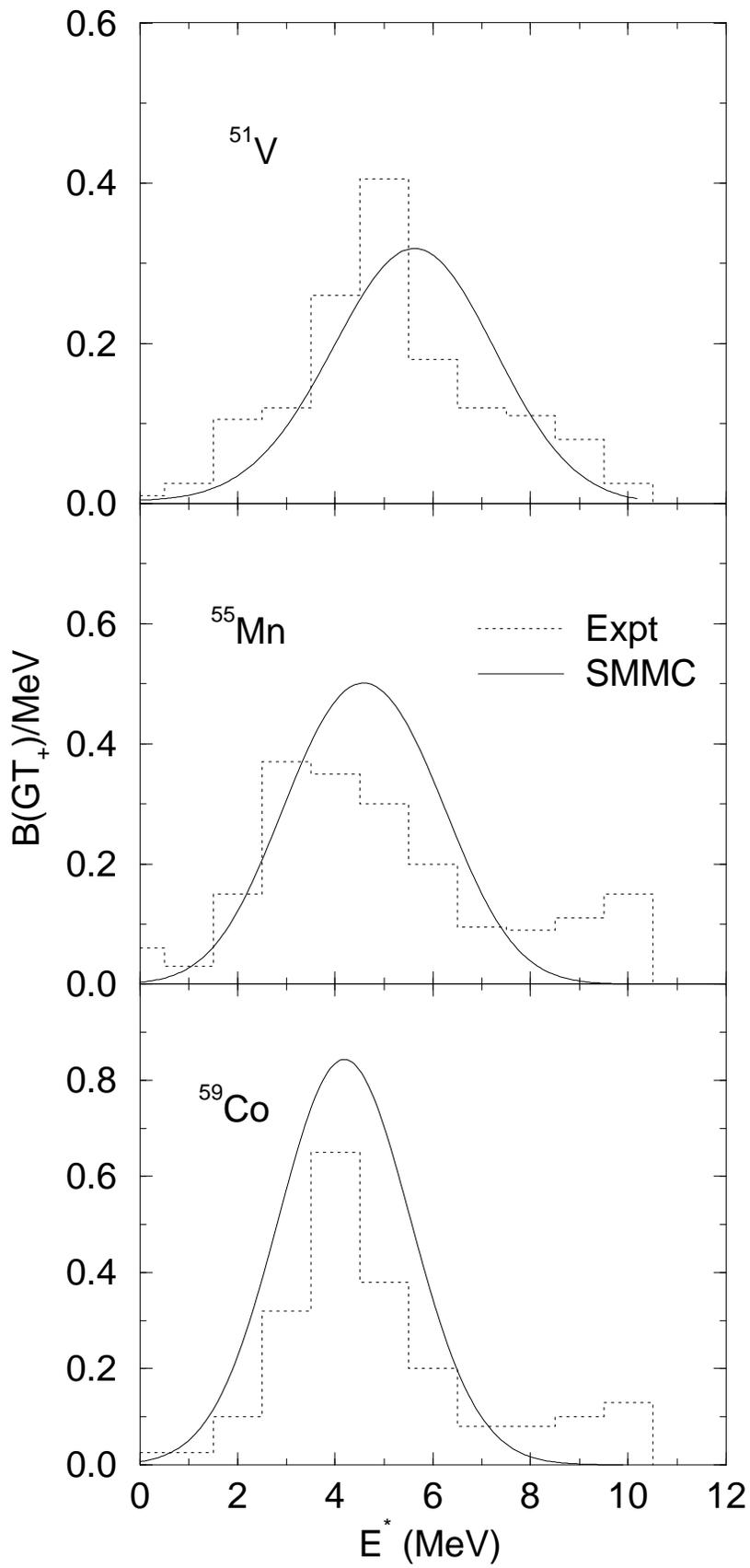
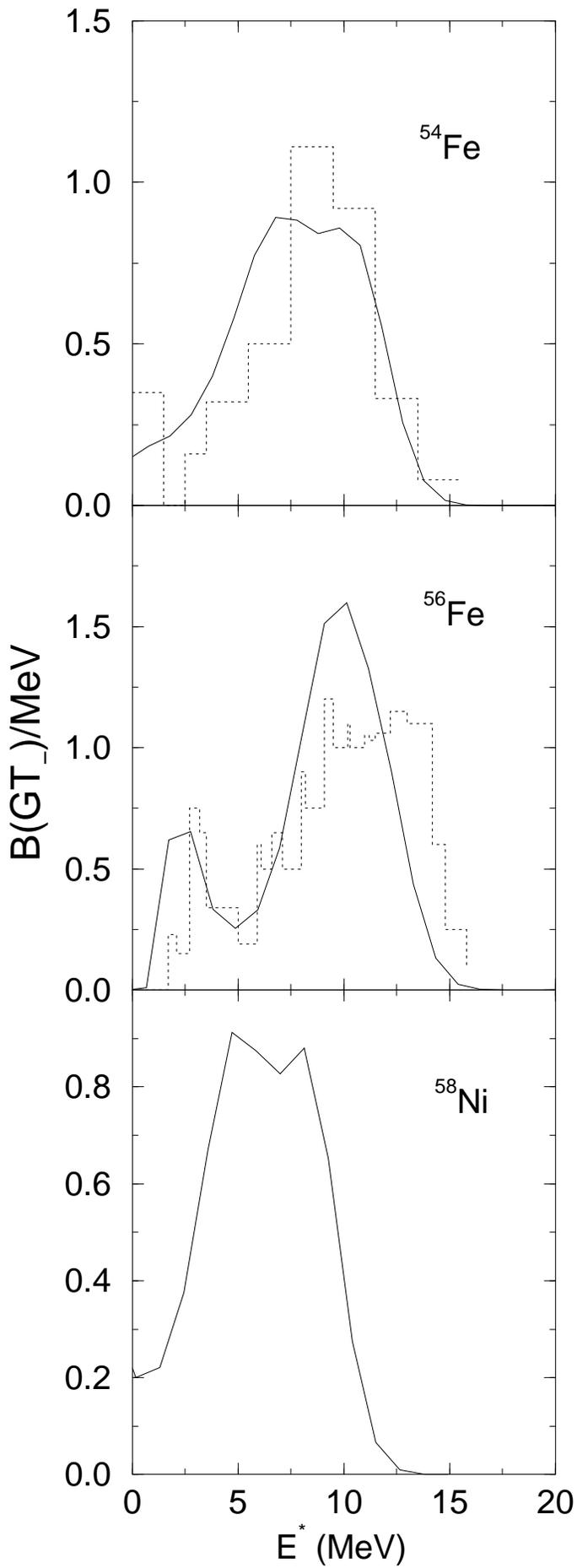
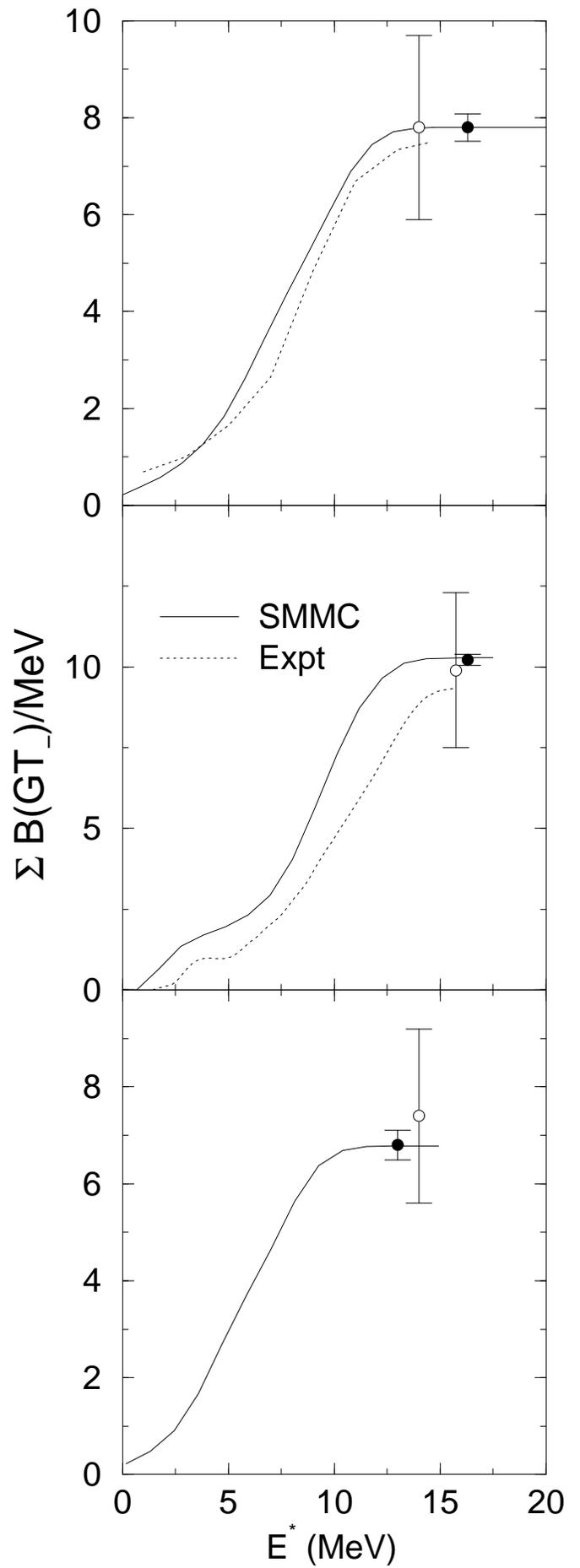
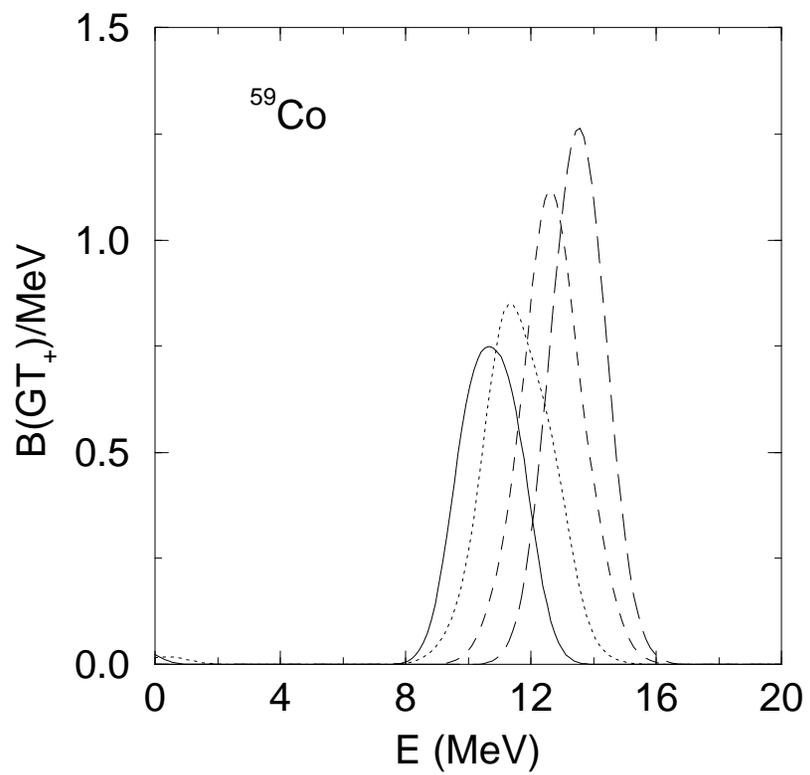
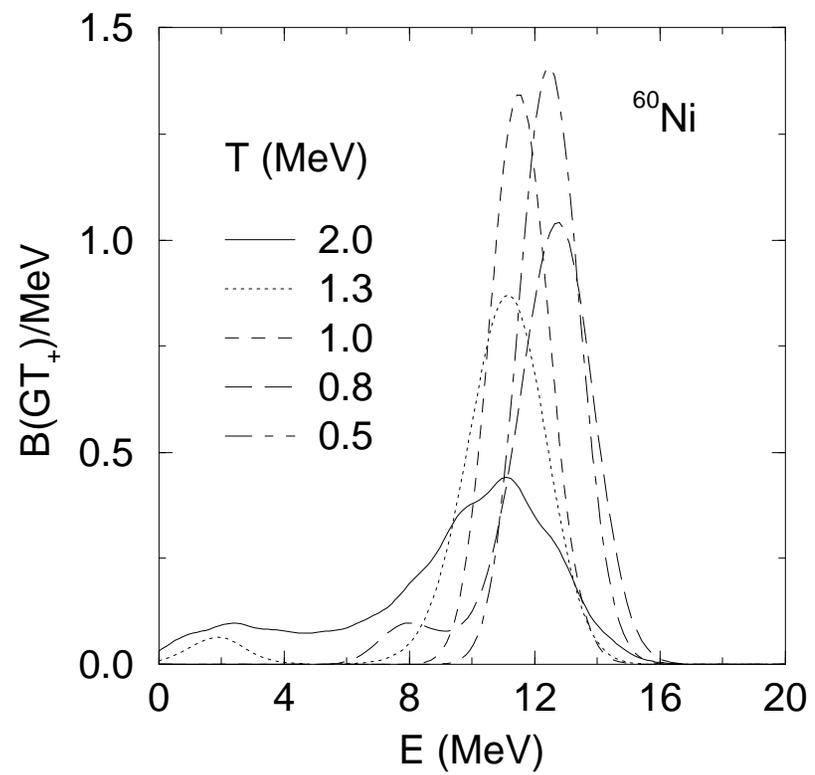
\end{document}